\documentclass[11pt]{article}
\input{psfig.sty}
\begin{document}

\title{
\textbf{Monte Carlo test of critical exponents and amplitudes
in 3D Ising and $\varphi^4$ lattice models}
}

\author{J. Kaupu\v{z}s
\thanks{E--mail: \texttt{kaupuzs@latnet.lv}} \\
Institute of Mathematics and Computer Science, University of Latvia\\
29 Rainja Boulevard, LV--1459 Riga, Latvia}

\date{\today}

\maketitle

\begin{abstract}
We have tested the leading correction--to--scaling exponent $\omega$
in \linebreak $O(n)$--symmetric models on a
three--dimensional
lattice by analysing the recent Monte Carlo (MC) data. We have found that
the effective critical exponent, estimated at finite sizes of the
system $L$ and $L/2$, decreases remarkably within the range of the
simulated $L$ values. This shows the incorrectness of some claims that
$\omega$ has a very accurate value $0.845(10)$ at $n=1$.
A selfconsistent infinite volume extrapolation yields
row estimates $\omega \approx 0.547$, $\omega \approx 0.573$, and
$\omega \approx 0.625$ at $n=1$, $2$, and $3$, respectively, in approximate
agreement with the corresponding exact values $1/2$, $5/9$, and $3/5$
predicted by our recently developed GFD (grouping of Feynman diagrams) theory.
We have fitted the MC data for the susceptibility of 3D Ising model at
criticality showing that the effective critical exponent $\eta$ tends
to increase well above the usually accepted values around $0.036$.
We have fitted the data within $[L;8L]$, including several terms in the
asymptotic expansion with fixed exponents, to obtain the effective
amplitudes depending on $L$. This method clearly demonstrates that
the critical exponents of GFD theory are correct
(the amplitudes converge to certain asymptotic values at $L \to \infty$),
whereas those of the perturbative renormalization group (RG) theory are
incorrect (the amplitudes diverge). A modification of the
standard Ising model by introducing suitable "improved" action (Hamiltonian)
does not solve the problem in favour of the perturbative RG theory.
\end{abstract}


\section{Introduction}

Since the exact solution of two--dimensional Ising model has
been found by Onsager~\cite{Onsager}, a study of various phase
transition models is of permanent interest. Nowadays, phase
transitions and critical phenomena is one of the most widely
investigated fields of physics. Remarkable progress has been
reached in exact solution of two--dimensional models~\cite{Baxter}.
Recently, we have proposed~\cite{K1} a novel method based on
grouping of Feynman diagrams (GFD) in $\varphi^4$ model.
Our GFD theory allows to analyse the asymptotic solution
for the two--point correlation function at and near criticality,
not cutting the perturbation series. As a result the possible
values of exact critical exponents have been obtained~\cite{K1} for
the Ginzburg--Landau ($\varphi^4$) model with $O(n)$ symmetry,
where $n=1, 2, 3, \ldots$ is the dimensionality of the order
parameter. Our predictions completely (exactly) agree with the
known exact and rigorous results in two dimensions~\cite{Baxter},
and are equally valid also in three dimensions. In~\cite{K1},
we have compared our results to some Monte Carlo (MC) simulations
and experiments~\cite{IS,SM,GA}. An additional comparison to MC data
has been made in~\cite{K2}. It has been shown~\cite{K1,K2} that the
actually discussed MC data for 3D Ising~\cite{IS,ADH} and
$XY$~\cite{SM} models are fully consistent with our theoretical
predictions, but not with those of the perturbative renormalization
group (RG) theory~\cite{Wilson,Ma,Justin}.
Some data for 3D Heisenberg model~\cite{Janke} also have been discussed
in~\cite{K2}. However, these data, likely, are not accurate enough and
here we reconsider the estimation of the critical point based on more
recent MC results. From the theoretical
(mathematical) point of view, the invalidity of the conventional
RG expansions has been demonstrated in~\cite{K1}.
The current paper, dealing with numerical analysis of
the three--dimensional $\lambda \varphi^4$ and Ising models, presents
one more confirmation that the correct values of critical exponents
are those predicted by the GFD theory.

\section{$\lambda \varphi^4$ model and its crossover to Ising model}

Here we discuss a $\varphi^4$ model
on a three--dimensional cubic lattice. The Hamiltonian of this model,
further called $\lambda \varphi^4$ model, is given by
\begin{equation} \label{eq:H}
H/T= \sum\limits_{\bf x} \left\{ -2 \kappa \sum\limits_{\mu}
\varphi_{\bf x} \varphi_{{\bf x}+\hat \mu} + \varphi_{\bf x}^2
+ \lambda \left( \varphi_{\bf x}^2 -1 \right)^2 \right\} \;,
\end{equation}
where the summation runs over all lattice sites,
$T$ is the temperature, \linebreak $\varphi_{\bf x} \in \, ]-\infty; +\infty[$
is the scalar order parameter at the site with coordinate ${\bf x}$,
$\hat \mu$ is a unit vector in the $\mu$--th direction,
$\kappa$ and $\lambda$ are coupling constants.
Obviously, the standard 3D Ising model is recovered in the limit
$\lambda \to \infty$ where $\varphi_{\bf x}^2$ fluctuations
are suppressed so that, for a relevant configuration,
$\varphi_{\bf x}^2 \simeq 1$ or $\varphi_{\bf x} \simeq \pm 1$ holds.
The MC data for the Binder cumulant in this
$\lambda \varphi^4$ model have been interpreted in accordance with
the $\epsilon$--expansion and a perfect agreement
with the conventional RG values of critical exponents has
been reported in~\cite{Hasenbusch}.
According to the definition in~\cite{Hasenbusch}, the Binder cumulant
$U$ is given by
\begin{equation} \label{eq:U}
U= \frac{\langle m^4 \rangle}{ \langle m^2 \rangle^2} \;,
\end{equation}
where
$m=L^{-3} \sum_{\bf x} \varphi_{\bf x}$ is the magnetization
and $L$ is the linear size of the system.
Based on the $\epsilon$--expansion, it has been
suggested in~\cite{Hasenbusch} that, in the thermodynamic limit
$L \to \infty$, the value of the Binder cumulant
at the critical point $\kappa=\kappa_c(\lambda)$ and, equally, at
a fixed ratio $Z_a/Z_p=0.5425$ (the precise value is not important)
of partition functions with periodic
and antiperiodic boundary conditions is a universal constant $U^*$
independent on $\lambda$. We suppose that the latter statement
is true, but not due to the $\epsilon$--expansion.
It is a consequence of some general argument of the RG theory:
on the one hand, $U$ is invariant under the RG transformation and,
on the other hand, an unique fixed point (not
necessarily the Wilson--Fisher fixed point) exists in the case of
an infinite system, so that $U \equiv U^*$ holds at $L \to \infty$
and $\kappa = \kappa_c(\lambda)$ where $U^*$ is the fixed--point
value of $U$. The above conclusion remains true if we allow
that the fixed point is defined not uniquely in the sense that it
contains some irrelevant degree(s) of freedom
(like $c^*$ and $\Lambda$ in the perturbative RG theory discussed
in Sec.~2 of~\cite{K1}) not changing $U$.
The numerical results in~\cite{HV} confirm the idea that
$\lim_{L \to \infty} U(L) = U^*$ holds at criticality, where $U^*$
is a universal constant independent on the specific microscopic
structure of the Hamiltonian.

\section{Estimation of the correction exponent $\omega$}
\label{sec:omega}

\begin{figure}
\centerline{\psfig{figure=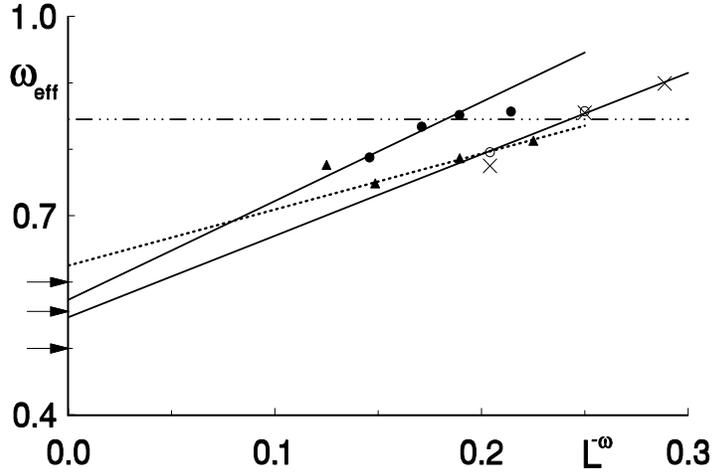,width=11cm,height=8.5cm}}
\vspace{-5ex}
\caption{\small Effective correction--to--scaling exponent
$\omega_{eff}(L)$ in the $O(n)$--symmetric $\lambda \varphi^4$ model
with $n=1$ (empty circles and crosses) and $n=3$ (triangles),
and $O(2)$--symmetric $dd-XY$ model (solid circles) depending on the
system size $L$. The linear least--squares fits give row estimates
of the asymptotic $\omega$ values $0.547$, $0.573$,
and $0.625$ at $n=1$, $2$, and $3$, respectively.
The corresponding theoretical values of the GFD theory
$1/2$, $5/9$, and $3/5$
(used in the $L^{-\omega}$ scale of the horizontal axis)
are indicated by arrows.
The dot--dot--dashed line shows the value $0.845(10)$ proposed
in~\cite{Hasenbusch} for the $3D$ Ising universality class ($n=1$).}
\label{om}
\end{figure}

\begin{figure}
\centerline{\psfig{figure=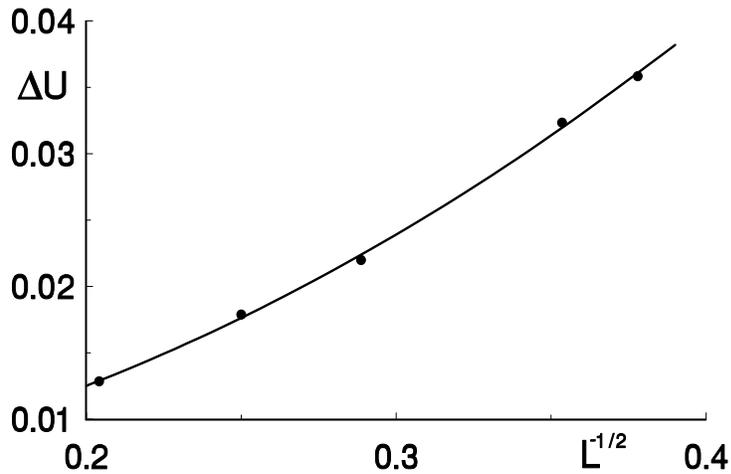,width=11cm,height=8.5cm}}
\vspace{-5ex}
\caption{\small The smoothened curve
$\Delta U(L)=0.003795-0.003232 L^{-1/2}+0.23433 L^{-1}$
for an approximation of $\Delta U(L)=U(L,0.8)-U(L,1.5)$ within
the interval $L \in [7;24]$.}
\label{du}
\end{figure}

Based on the idea that $U^*$ is constant for a given universality
class, here we estimate the correction--to--scaling exponent $\omega$.
According to~\cite{K2}, corrections to finite--size scaling
for the magnetization of the actual 3D Ising and
$\lambda \varphi^4$ models are represented by an expansion in terms
of $L^{-\omega}$ where $\omega=1/2$. One expects that the
magnetization (Binder) cumulant~(\ref{eq:U}) has the same singular
structure. Since $\lim_{L \to \infty} U(L,\lambda) \equiv U^*$ holds at
a fixed ratio $Z_a/Z_p$, a suitable ansatz for estimation
of $\omega$ is~\cite{Hasenbusch}
\begin{equation} \label{omef}
U(L,\lambda_1)-U(L,\lambda_2) \simeq const \cdot L^{-\omega}
\hspace{4ex} \mbox{at} \hspace{2ex} Z_a/Z_p=0.5425 \;,
\end{equation}
which is valid for any two different nonzero values $\lambda_1$
and $\lambda_2$ of the coupling constant $\lambda$.
The data for $\Delta U(L)= U(L,0.8)-U(L,1.5)$ can be read
from Fig.~1 in~\cite{Hasenbusch} (after a proper magnification)
without an essential loss of the numerical accuracy, i.~e., within
the shown error bars. Doing so, we have evaluated the effective exponent
\begin{equation}
\omega_{eff}(L) = \ln \left[ \Delta U(L/2)/ \Delta U(L) \right] / \ln 2 \;,
\end{equation}
i.~e., $\omega_{eff}(12) \simeq 0.899$,
$\omega_{eff}(16) \simeq 0.855$, and $\omega_{eff}(24) \simeq 0.775$.
These values are shown in Fig.~\ref{om} by crosses.
Such an estimation,
however, can be remarkably influenced by the random scattering
of the simulated data points, particularly, at larger sizes where
$\Delta U(L)$ becomes small. This effect can be diminished if the
values of $\Delta U(L)$ are read from a suitable smoothened curve.
A comparison to the original results (without the smoothening)
provides some objective criterion of the accuracy of such
estimations. We have found that $\Delta U(L)$ within $L \in [7;24]$
can be well approximated by a second--order polinomial in $L^{-1/2}$,
as shown in Fig.~\ref{du}.
Without any claims about validity of such
an approximation well outside of this interval, we can consider
the least--squres fit in Fig.~\ref{du} as an appropriate smoothened
curve from which we read $\omega_{eff}(16) \simeq 0.8573$ and
$\omega_{eff}(24) \simeq 0.7956$. These values are depicted in
Fig.~\ref{om} by empty circles. As we see, the results are similar
to those obtained by a direct calculation from the original
data points (crosses). However, the values obtained from the
smoothened curve (circles) are more
accurate and reliable. As regards the smallest size, we suppose
that the original estimate $\omega_{eff}(12) \simeq 0.899$ is
accurate enough even without any smoothening, since the values
of $\Delta U(6)$ and $\Delta U(12)$ are large relative to the
statistical errors.

In such a way, we see from Fig.~\ref{om}
that the effective exponent $\omega_{eff}(L)$ decreases
remarkably with increasing of $L$. According to GFD theory,
$\omega_{eff}(L)$ is a linear function of $L^{-1/2}$ at $L \to \infty$,
as consistent with the expansion in terms of $L^{-\omega}$ where
$\omega=0.5$. More data points,
including larger sizes $L$, are necessary for a reliable
estimation of the asymptotic exponent
$\omega =\lim_{L \to \infty} \omega_{eff}(L)$. Nevertheless, already
a row linear extrapolation in the scale of $L^{-1/2}$ 
with the existing data points yields the result $\omega \approx 0.547$
which is reasonably close to the exact value $0.5$ (horizontal dashed
line in Fig.~\ref{om}) found within
the GFD theory. The corresponding least--squares fit with circles
(at $L=24,16$) and cross (at $L=12$) is shown in Fig.~\ref{om} by
a straight solid line. It is evident from Fig.~\ref{om} that the final
result $\omega=0.845(10)$ (horizontal dot--dot--dashed line) reported
in~\cite{Hasenbusch} represents some average
effective exponent for the interval $L \in [6;24]$.
It has been claimed in~\cite{Hasenbusch} that the estimates for
$\omega$ (cf. Tab.~2 in~\cite{Hasenbusch}) are rather stable
with respect to $L_{min}$, where $L_{min}$ is the minimal lattice
size used in the fit. Unfortunately, the analysis
has been made in an obscure fashion, i.~e., giving no original
data, so that we cannot check the correctness of this claim.
Besides, the estimates in Tab.~2 of~\cite{Hasenbusch} has been made
by using an ansatz
\begin{equation} \label{omef1}
U(L,\lambda) = U^* + c_1(\lambda) L^{-\omega}
\hspace{4ex} \mbox{at} \hspace{2ex} Z_a/Z_p=0.5425 \;,
\end{equation}
which is worse than~(\ref{omef}). Namely, (\ref{omef}) and (\ref{omef1})
are approximations of the same order, but~(\ref{omef1}) 
contains an additional parameter $U^*$ which is not known precisely. 
The results of an analysis with the ansatz~(\ref{omef}), reflected
in Tab.~5 of~\cite{Hasenbusch}, are not convincing, since
only very small values of $L_{min}$ (up to $L_{min}=6$) have been
considered.

In any case, we prefer to rely on that information we can
check, and it shows that the claim
in~\cite{Hasenbusch} that $\omega=0.845(10)$ holds with
$\pm 0.01$ accuracy cannot be correct,
since $\omega_{eff}(L)$ is varied in the first decimal place.

 We have made a similar estimation of $\omega$ for $O(n)$--symmetric
spin models, namely, for the dynamically diluted $XY$ ($dd-XY$)
model simulated in~\cite{CHPRV} ($n=2$) and for $O(3)$--symmetric
$\lambda \varphi^4$ model simulated in~\cite{Has3}. In the case of
the $dd-XY$ model, parameter $D$ (cf.~Eq.(6) in~\cite{CHPRV})
plays the role of $\lambda$ in~(\ref{omef}). The data for the Binder
cumulant in Fig.~1 of~\cite{CHPRV} look rather accurate, i.~e., not
scattered. This enables us to estimate $\omega_{eff}$ just from
the data at $D=1.03$ and $D=\infty$ ($XY$ model).
The resulting values of $\omega_{eff}$
are depicted in Fig.~\ref{om} by solid circles. The scale of
$L^{-\omega}$ is used, where $\omega=5/9$ is our
theoretical value of the correction--to--scaling exponent at $n=2$
consistent with the general hypothesis proposed in~\cite{K2}.
As we see, the solid circles can be well located on a smooth line
which, however, is remarkably curved at smaller sizes. Due to the
latter reason, we have used only the last three points (the largest
sizes) for the linear fit (solid line) resulting in an 
estimate $\omega \approx 0.573$ which comes close to our theoretical
value $\omega=5/9=0.555 \ldots$

\begin{figure}
\centerline{\psfig{figure=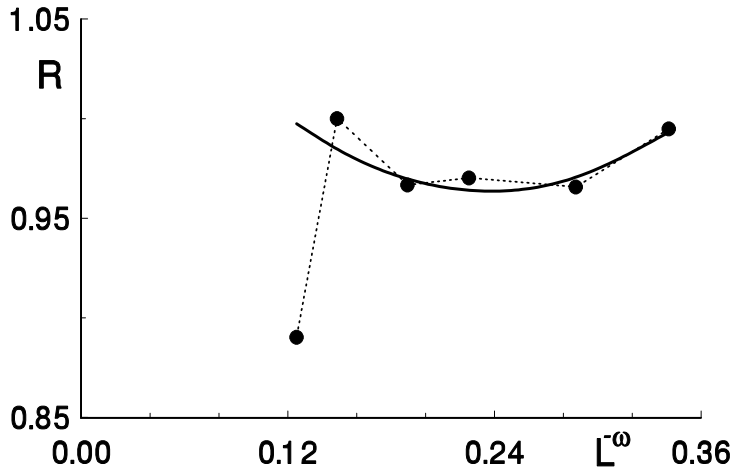,width=11cm,height=8.5cm}}
\vspace{-5ex}
\caption{\small The ratio $R(L)=(U(L,2)-U^*)/(U^*-U(L,\infty))$
for $O(3)$--symmetric $\lambda \varphi^4$ model estimated from
the MC data of~\cite{Has3}. The scale of $L^{-\omega}$ with
$\omega=0.6$ is used. The solid line (parabola) represents the
least--squares fit including only four smallest sizes.}
\label{r}
\end{figure}

 The estimates of $\omega_{eff}$ for the $O(3)$-symmetric
$\lambda \varphi^4$ model are depicted in Fig.~\ref{om}
(in the scale of $L^{-\omega}$ with our $\omega$ value $0.6$)
by triangles. The data have been extracted from Fig.~1
of~\cite{Has3} at $\lambda_1=2$ and $\lambda_2=\infty$.
The obtained $\omega_{eff}(L)$ values at $L=12$, $16$, and $24$
well lie on a straight line (tiny--dashed line in
Fig.~\ref{om}) yielding an asymptotic estimate $\omega \approx 0.625$
which is reasonably close to our theoretical prediction $\omega=0.6$.
Hence, the $\omega_{eff}(32)$ value deviates upwards unexpectedly.
We suppose, this is due to an inaccurate simulation of the largest
size $L=32$, as explained below. We have depicted in Fig.~\ref{r}
the ratio $R(L)=(U(L,2)-U^*)/(U^*-U(L,\infty))$ evaluated from
the data in Fig.~1 of~\cite{Has3} with $U^*=1.14022$ (the average
over three estimates at a fixed $Z_a/Z_p$ given in Tab.~2 in~\cite{Has3}).
According to~(\ref{omef1}), $R(L)$ tends to a constant at $L \to \infty$.
The $R(24)$ data point well lie on the smooth line (parabola) representing
the least--squares fit to four smallest sizes $L=6, 8, 12$, and $16$,
whereas the simulated $R(32)$ value drops down unreasonably.
So, it looks like a wrong simulation has been made at $L=32$ to
confirm the known RG estimate $\omega \approx 0.8$.
Thus, our extrapolation in Fig.~\ref{om}
(tiny--dashed line), omiting the point with $L=32$, is justified.

In summary, the extrapolated $\omega$ values (Fig.~\ref{om})
in all three cases $n=1,2,3$ are reasonably close to our theoretical
values $1/2$, $5/9$, and $3/5$ indicated by arrows. Only a small
systematic deviation is observed. This, likely, is due to the error of
linear extrapolation: the $\omega_{eff}(L)$ plots have a
tendency to curve down slightly.
The conventional (RG) estimate $\omega \approx 0.8$ more or less
corresponds to effective exponents for currently simulated finite
system sizes, but not to the asymptotic exponents.

The data for $n=4$ also are available in~\cite{Has3}. Unfortunately,
they are too much scattered for the actual analysis.

\section{The critical coupling of 3D Ising \\ and Heisenberg models}
\label{sec:crp}

 A conventional method to determine the critical exponent $\eta$
is a fit of the susceptibility data at criticality. For this, however,
we need a very accurate value of the critical coupling (temperature).
In this section we discuss the estimation of the critical coupling
$\beta_c$ for 3D Ising and Heisenberg models.

The critical point of the standard 3D Ising model has been estimated
in~\cite{HV} with a $7$--digit accuracy, i.~e. $\beta_c=0.2216545$. We have
made our own fits with the MC data of~\cite{HV} to check the accuracy
of this estimation, and have obtained the same value within
error bars of $10^{-7}$. We will use in our further analysis also a
similar estimate $\beta_c=0.383245$~\cite{HV} for the so called
"improved" 3D Ising model.

The critical coupling of the classical 3D Heisenberg model
is known much less accurately than that of the 3D Ising model.
Some of the known estimates are $\beta_c=0.6929(1)$~\cite{PFL},
$\beta_c=0.693035(37)$~\cite{CFL}, and
$\beta_c=0.693001(10)$~\cite{Ballesteros}.

\begin{figure}
\centerline{\psfig{figure=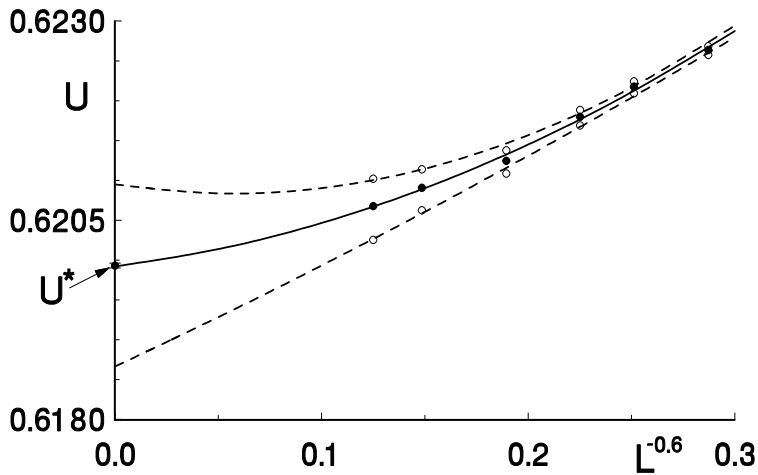,width=11cm,height=8.5cm}}
\vspace{-5ex}
\caption{\small The Binder cumulant $U$ vs system size $L$ in the
3D Heisenberg model at $\beta=0.6929$ (bottom), $\beta=0.692955$
(middle), and $\beta=0.693$ (top). Symbols depict the
MC data of~\cite{CB}, whereas lines represent the least--squares
approximations of these data by a parabola. The solid line,
coinciding with the universal critical value of the Binder cumulant
$U^*=0.61993(3)$, corresponds to
the critical coupling $\beta_c \simeq 0.692955$.}
\label{crp}
\end{figure}

 In principle, the location of the critical point can be found with
a high accuracy and reliability by simulations of the Binder
cumulant $U(\beta,L)$ in close vicinity of the critical point, as it
has been done in~\cite{HV} for the 3D Ising model. Taking into account
the leading and the subleading corrections to scaling, we have
\begin{equation} \label{Ubc}
U(\beta_c,L) \simeq U^* +a_1 L^{-\omega} +a_2 L^{-2\omega} \;.
\end{equation}
If the universal value of $U^*$ is known with a high precision,
then the critical value of $\beta$ at which $U(\beta,L)$ coincides
with~(\ref{Ubc}) is well defined. Namely, at $\beta=\beta_c$ the
quadratic (least--squares) extrapolation of $U(\beta,L)$ in the scale of
$L^{-\omega}$ should yield $U^*$ at $L^{-\omega} \to 0$.
In Fig.~\ref{crp}, we have shown the results of such an extrapolation
with $\omega=0.6$ (our theoretical value)
at three different values of $\beta$, i.~e., $0.6929$ (lower dashed line),
$0.692955$ (solid line), and $0.693$ (upper dashed line). The data for
$U(\beta,L)$ have been extracted from Fig.~1 in~\cite{CB} via an
approximation
$U(\beta,L) \simeq U(\beta_0,L) + 0.06 \, L^{1.4} (\beta-\beta_0)$,
where $\beta_0=1/1.4432$.
Note that, in distinction to~(\ref{eq:U}),
now we use the conventional definition
$U=1- (1/3) \langle m^4 \rangle / \langle m^2 \rangle^2$.
A suitable estimate of $U^*$, taken from Tab.~2 in~\cite{Has3}, is then
$U^*=0.61993(3)$. This value with the error bars is indicated in
Fig.~\ref{crp} by an arrow.
As we see from Fig.~\ref{crp}, the estimate $\beta_c \simeq 0.692955$ is
consistent with this value of $U^*$. Our estimation is rather stable, i.~e.,
if the extrapolation is made in the scale of $L^{-0.8}$, we get
practically the same result $\beta_c \simeq 0.692957$. Taking into account
the curvature of the $U(\beta,L)$ plot at $\beta=0.693$, it is unlikely
that $0.693$ could be the correct value of $\beta_c$ yielding $U^*=0.61993(3)$
at $L \to \infty$. According to the strong variation of the extrapolated
$U^*$ value with $\beta$, it is plausible that the error of our estimation
$\beta_c \simeq 0.692955$ is about $0.00001$ or even smaller.

Our result agree within the error bars with the value
$0.6929(1)$ of~\cite{PFL}, while the error bars of the estimates
$\beta_c=0.693035(37)$~\cite{CFL} and
$\beta_c=0.693001(10)$~\cite{Ballesteros}, in our opinion, are
underestimated. As regards the value of~\cite{CFL}, this is a result
of a linear extrapolation (in the scale of $L^{-1/0.7036}$) of temperature
values corresponding to extrema points for several physical quantities.
However, the shift in $\beta$ as large as $0.0001$ is practically
invisible in the scale of Fig.~3 in~\cite{CFL}. If one allows that
the lines in this figure are curved very slightly, then even larger
deviation in the extrapolated $\beta_c$ value is possible. In other
words, the proposed error bars $\pm 0.000037$, obviously, include
only the statistical error and are underestimated since they ignore
the possible systematic deviation due to the neglected corrections to scaling.
A similar problem with the subleading correction persists
in~\cite{Ballesteros}. Namely, a linear extrapolation in Fig.~\ref{crp}
would give a larger $\beta_c$ value, close to $0.693$ (in agreement with
that of~\cite{Ballesteros}), but this value is shifted down to $0.692955$
due to the subleading correction in the asymptotic expansion of $U$.
This correction has been neglected in the analysis of the
Binder cumulant crossings in~\cite{Ballesteros}.
 We have taken into account both the leading and the subleading
corrections to scaling and, therefore, our value $0.692955$ is more
accurate than those proposed in~\cite{CFL} and~\cite{Ballesteros}, unless 
the data of~\cite{CB} contain large systematical errors.

\section{Fitting the susceptibility data at criticality}
\label{sec:fit}

In this section we discuss some fits of MC data at criticality.
According to the finite--size scaling theory, the susceptibility
$\chi$ near the critical point is represented by an expansion
\begin{equation} \label{chi}
\chi= L^{2-\eta} \left( g_0(L/\xi)
+ \sum\limits_{l \ge 1} L^{-\omega_l} g_l(L/\xi) \right) \;,
\end{equation}
where $g_l(L/\xi)$ are the scaling functions, $\xi$ is the correlation
length of an infinite system, $\eta$ is the critical exponent
related to the $k^{-2+\eta}$ divergence of the correlation function
in the wave vector space at criticality, and $\omega_l$ are
correction--to--scaling exponents, $\omega_1 \equiv \omega$
being the leading correction exponent. The correlation length
diverges like $\xi \propto t^{-\nu}$ at $t \to 0$, where
$t=1-\beta/\beta_c$ is the reduced temperature. Thus,
for large $L$, in close vicinity of the critical point
where $tL^{1/\nu} \ll 1$ holds Eq.~(\ref{chi}) can be
written as
\begin{equation} \label{chi1}
\chi= a \, L^{2-\eta} \left( 1 + \sum\limits_{l \ge 1} b_l
L^{-\omega_l} + \delta(t,L) \right) \;,
\end{equation}
where $a=g_0(0)$ and $b_l=g_l(0)/g_0(0)$ are the amplitudes, and
$\delta(t,L)$ is a correction term which takes into account the
deviation from criticality. In the first approximation it reads
\begin{equation} \label{delta}
\delta(t,L) \simeq c \cdot tL^{1/\nu} \;,
\end{equation}
where $c$ is a constant.

\begin{figure}
\centerline{\psfig{figure=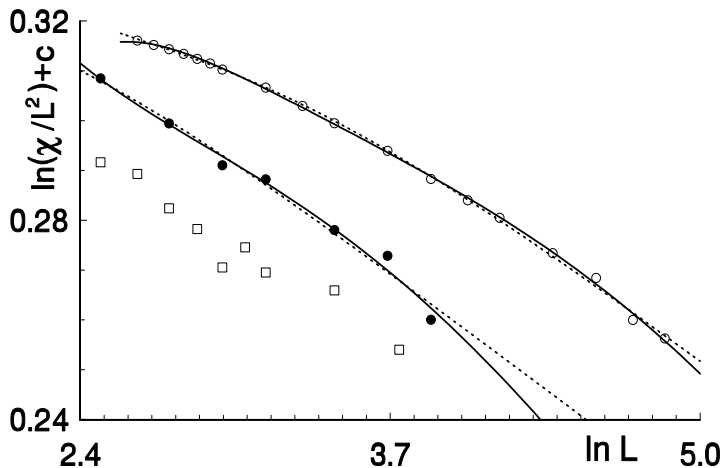,width=11cm,height=8.5cm}}
\vspace{-5ex}
\caption{\small The fits of $\ln \left( \chi/L^2 \right)$ data
at criticality (ansatz~(\ref{chi1})) shifted by a constant $c$.
Circles represent the MC data for 3D Ising model~\cite{HV} at
$\beta=0.2216545$ ($c=0$, empty symbols) and 3D Heisenberg
model~\cite{Janke} at $\beta=0.692955$ ($c=0.55$, solid symbols).
The corresponding fits with our (GFD) exponents 
($\ln a=1.065289, b_1=-2.72056, b_2=8.18636, b_3=-10.49614$
and $\ln a=0.207324, b_1=-1.22546, b_2=1.85823$) are shown
by solid lines, whereas those with the exponents
of~\cite{Hasenbusch,Justin1}
($\ln a=0.430933, b_1=0.05850, b_2=-7.74767, b_3=12.42890$ and
$\ln a=-0.150242, b_1=0.03947, b_2=-0.45033$) -- by
tiny--dashed lines.
The empty boxes are MC data for 3--component 3D $XY$ model~\cite{NhM},
shifted by $c=0.85$.} 
\label{chifi}
\end{figure}

 We start our analysis with the standard 3D Ising model with the Hamiltonian
\begin{equation}
H/T= -\beta \sum\limits_{\langle i j \rangle} \sigma_i \sigma_j \;.
\end{equation}
The critical point of this model with a 7--digit accuracy is
$\beta_c \simeq 0.2216545$ (Sec.~\ref{sec:crp}).
From the maximal values of the derivative
$\partial \ln \langle m^2 \rangle / \partial \beta \equiv
\partial \ln \chi / \partial \beta$ evaluated in~\cite{FL} we conclude
that the shift of $\beta$ by $10^{-7}$ produces
the variation of $\ln \chi$ at $L=96$ near
$\beta=\beta_c$, which does not exceed $4.7 \cdot 10^{-4}$ in magnitude.
The latter means that, with a good enough accuracy, we may assume
that $\beta_c$ is just $0.2216545$ when fitting the susceptibility
data at criticality within $L \in [4;128]$. Here we mean the MC data
given in Tab.~25 of~\cite{HV}.
We have made and compared several fits of these data to 
ansatz~(\ref{chi1}) with $\delta(t,L)=0$
(more precisely, to the corresponding formula
for $\ln \chi$) for two different sets of the critical
exponents, i.~e., our (GFD) and that proposed in~\cite{Hasenbusch}.
The fits made with our exponents systematically
improve relative to those made with the exponents of~\cite{Hasenbusch},
as the system sizes grow and the approximation order increases.
The necessity to include several correction terms is dictated
by the fact that corrections to scaling are rather strong.
According to the least--squares criterion, the fit with our exponents
$\eta=1/8$ and $\omega_l=l/2$ becomes better than that provided
by the more conventional exponents $\eta=0.0358(4)$, $\omega_1=0.845(1)$,
$\omega_2=2 \omega_1$, and $\omega_3=2$~\cite{Hasenbusch} starting with
$L_{min}=28$ (i.~e., $L \in [L_{min};128]$), if two correction terms
($l=1,2$) are included.
In the case of three correction terms it occurs already at $L_{min}=11$.
The four--parameter ($a$, $b_1$, $b_2$, $b_3$) fits to MC data (empty
circles) within $L \in [14;128]$ are shown in Fig.~\ref{chifi}.
The fit with our exponents (upper solid line) is relatively better at
larger sizes. However, both fits (upper solid and dashed lines) look,
in fact, quite similar, so that we cannot make unambiguous conclusions
herefrom.

We have shown in Fig.~\ref{chifi} also the three--parameter fits to
the susceptibility
data of the 3D Heisenberg model at $\beta=\beta_c \simeq 0.692955$
(Sec.~\ref{sec:crp}) extracted from Fig.~6 in~\cite{Janke} by
a suitable linear interpolation. The MC data within $L \in [12;48]$
are shown by solid circles. The fit with our exponents
($\eta=0.1$, $\omega_l= 0.6 \, l$)
is depicted by lower solid line, whereas that with the conventional
RG exponents ($\eta=0.0355$, $\omega=0.782$~\cite{Justin1}) -- by
lower tiny--dashed line. Like in the case of 3D Ising model, the fit
with our exponents looks slightly better, although the  MC data
are too inaccurate to make serious conclusions herefrom.
Comparing the amplitudes given in the caption of Fig.~\ref{chifi},
we see that corrections to scaling are remarkably weaker in
the 3D Heisenberg model as compared to the 3D Ising model, so that
the neglected third--order correction in the case of the Heisenberg model
could be small enough even at sizes somewhat below $L=12$.
In this aspect, it is interesting to mention that the fit with
our exponents $\eta=0.1$ and $\omega_l= 0.6 \, l$ (and not the other one)
qualitatively correctly reproduces the shape of the actual
$\ln \left( \chi/L^2 \right)$ plot at $L<12$ where it curves upwards
to meet the condition $\chi(L=1)= \langle \sigma^2 \rangle =1$.

\begin{figure}
\centerline{\psfig{figure=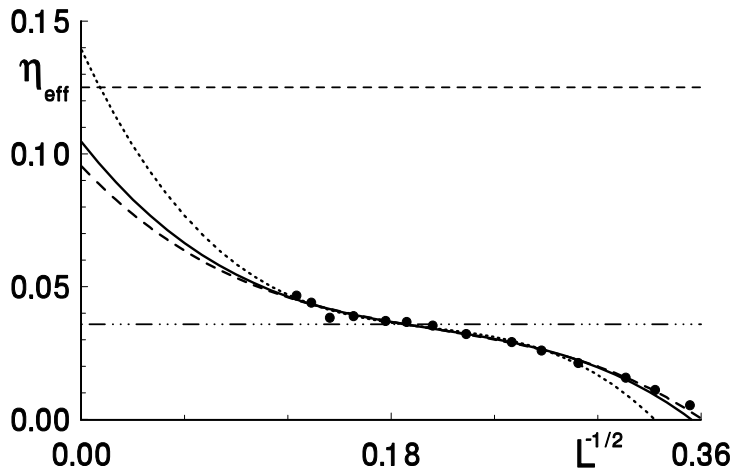,width=11cm,height=8.5cm}}
\vspace{-5ex}
\caption{\small The effective critical exponent $\eta_{eff}(L)$ (solid
circles) obtained by fitting the susceptibility data of 3D Ising model
at criticality ($\beta=0.2216545$)~\cite{HV} within the interval $[L;2L]$.
The least--squares approximations obtained by fitting the $\eta_{eff}(L)$
data within $[L_{min};64]$ to a third--order polinomial in $L^{-1/2}$ are
shown by dashed ($L_{min}=9$), solid ($L_{min}=10$), and tiny--dashed
($L_{min}=12$) lines. The asymptotic value $\eta=1/8$ of the GFD
theory is indicated by a horizontal dashed line. The dot--dot--dashed
line represents the $\eta$ value $0.0358$ proposed in~\cite{Hasenbusch}.}
\label{eteff}
\end{figure}

For comparison, we have shown in Fig.~\ref{chifi} also the MC data
for 3D $XY$ model~\cite{NhM} in which, however, only $x$ and $y$ components
of the 3--component order parameter interact with each other. One
believes~\cite{NhM} that this model belongs to the universality class of
the standard $XY$ model with the number of components $n=2$. Unfortunately,
we have not found in the recent literature more accurate explicit data
for $n=2$ case. As we see, the actual MC data (empty boxes) at
$\beta_c$ evaluated approximately $\beta_c \simeq 0.6444$~\cite{NhM} are
rather scattered and, therefore, unsuitable for a refined analysis.
Nevertheless, this is a typical situation where authors of such data make
a very "accurate" and "convincing" estimation $\gamma/\nu=1.9696(37)$
or $\eta=0.0304(37)$ making a simple linear fit. However, the refined
analysis given above has shown that even in the case of 3D Ising model,
where the data are incompatibly more accurate, it is not so easy to
distinguish between $\eta=0.0358$ and $\eta=1/8$. Moreover, a refined
analysis prefer the second value which is much larger than those
usually provided by linear fits at typical system syzes $L \le 48$.
This is particularly well seen in Fig.~\ref{eteff}, where
the effective critical exponent $\eta_{eff}(L)$ of the
3D Ising model, estimated via the linear fit within $[L;2L]$, is depicted
by solid circles.
As we see, $\eta_{eff}(L)$ tends to increase well above
the conventional value $0.0358$ (horizontal dot--dot--dashed line).
The shape of the $\eta_{eff}(L)$ plot is satisfactory well reproduced
by a third--order polinomial in the actual scale of $L^{-1/2}$.
Three such kind of least--squares approximations
(at $L_{min}=9,10,12$) are shown in Fig.~\ref{eteff}.
These fits do not provide very accurate
and stable asymptotic values of $\eta$. Nevertheless, they are
more or less in agreement with our theoretical prediction
$\eta=1/8$ (horizontal dashed line). Besides, the values of $\eta_{eff}$
are affected by the error in $\beta_c$ (about $10^{-7}$) only slightly,
i.~e., by an amount not exceeding $0.001$.

\section{A test for 3D Ising model with "improved" action}
\label{imp}

Here we discuss some estimations of the critical exponents
from the susceptibility data of 3D Ising model, reported in~\cite{HV},
with the so called "improved" action (i.~e., $H/T$).
One of the problems with the standard 3D Ising model is that corrections
to scaling are strong. It has been proposed in~\cite{HV} to solve
this problem by considering a modified (spin--1) Ising model
with the Hamiltonian
\begin{equation} \label{Hsp1}
H/T= - \beta \sum\limits_{\langle i j \rangle}
\sigma_i \sigma_j + D \sum\limits_{i} \sigma_i^2  \;,
\end{equation}
where the spin $\sigma_i$ takes the values $0, \pm 1$,
with two coupling constants $\beta$ and $D$ adjusted in such a way that
the leading correction to
finite--size scaling vanishes for all relevant physical quantities
(magnetization cumulant, energy per site, susceptibility, etc.) and their
derivatives. Moreover, according to the claims in~\cite{HV}
(see the conclusions in~\cite{HV}), the ratios of
the leading and subleading corrections are universal, so that not only
the leading but all (!) corrections should vanish simultaneously.

\begin{figure}
\centerline{\psfig{figure=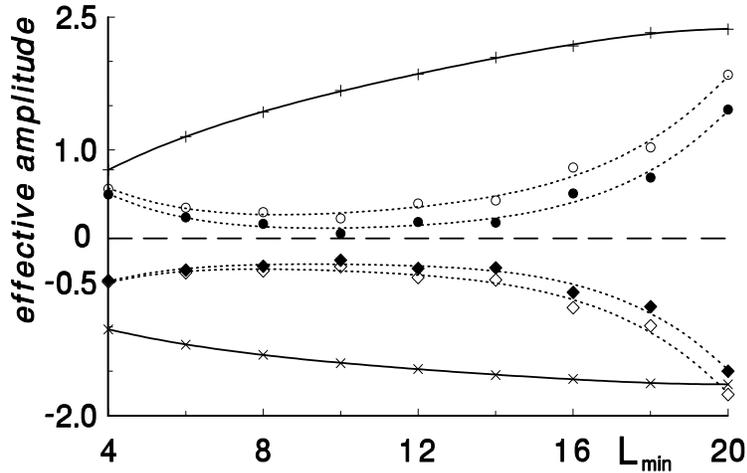,width=11cm,height=8.5cm}}
\vspace{-5ex}
\caption{\small The effective amplitudes $10 b_1$ (circles) and
$b_2$ (rhombs) in~(\ref{chi1}) estimated at fixed exponents
$\eta=0.0358$, $\omega_1=0.845$, $\omega_2=2 \omega_1$, and
$\nu=0.6296$ by fitting the MC data within $L \in [L_{min};56]$.
Filled symbols correspond to $\delta(t,L)=0$, empty
symbols -- to $\delta(t,L)=10^{-6} L^{1/\nu}$. The effective
amplitudes $b_1$ and $b_2$ estimated with the critical
exponents of our GFD theory ($\eta=1/8$, $\omega_l=l/2$) at
$\delta(t,L)=0$ are shown by "x" and "+", respectively.
Lines represent the least--squares approximations by a fourth--order
polinomial in $L$.}
\label{b}
\end{figure}

We have checked the correctness of these claims as described below.
We have fitted the corresponding to~(\ref{chi1}) expression for
$\ln \chi$ to the susceptibility data of the "improved" 3D Ising
model~(\ref{Hsp1}) with $(\beta,D)=(0.383245, 0.624235)$
(this is an approximation of the critical point) given
in~\cite{HV} (Tab.~26). By fixing the exponents, the least--squares
fit within $L \in [L_{min};56]$ (here $L=56$ is the maximal size
available in Tab.~26 of~\cite{HV}), including the leading and the
subleading correction to scaling, provides the effective amplitudes
$a$, $b_1$, and $b_2$ depending on $L_{min}$.
We have made a test with the critical exponents $\eta=0.0358(4)$,
$\omega=0.845(10)$, and $\nu=0.6296(3)$ proposed in~\cite{Hasenbusch}.
These values are close to those of the usual RG expansions~\cite{Justin1},
but, as claimed in~\cite{Hasenbusch}, they are more accurate.
According to~\cite{Hasenbusch}, the asymptotic expansion contains corrections
like $L^{-n \omega}$ and $L^{-2n}$, where $n=1, 2, 3, \ldots$
Thus we have $\omega_1=\omega$ and $\omega_2=2 \omega$.
The resulting amplitudes $10 b_1(L_{min})$ and $b_2(L_{min})$
are shown in Fig.~\ref{b} by circles and rhombs, respectively.
We have depicted by filled symbols
the results of the fitting with $\delta(t,L)=0$, assuming that the
critical coupling $\beta_c=0.383245$ has been estimated in~\cite{Hasenbusch}
with a high enough (6 digit) accuracy. The data points quite
well fit smooth (tiny dashed) lines within $L_{min} \in [4;20]$, which
means that the statistical errors are reasonably small.
If the exponents used in the fit are correct and corrections to
scaling are small indeed, then the convergence of the effective amplitudes
to some small values is expected with increasing of $L_{min}$.
However, as we see from Fig.~\ref{b}, the effective amplitudes tend
to increase in magnitude acceleratedly as $L_{min}$ exceeds $14$.
A small inaccuracy in $\beta_c$ value can be compensated by
the term $\delta(t,L) \simeq c^* L^{1/\nu}$ in~(\ref{chi1}),
where $c^*=ct$ (cf.~Eq.~(\ref{delta})). The results of fitting with
$c^*=10^{-6}$ are shown in Fig.~\ref{b} by empty symbols. As we see,
the expected inaccuracy in $\beta_c$ of order $10^{-6}$ does not
change the qualitative picture. The increase of the effective
amplitudes indicates that either the exponents are false, or
the asymptotic amplitudes are not small (or both). This is our
argument that the claims in~\cite{HV} about very
accurate critical exponents, extracted from the 3D Ising model
with "improved" action, are incorrect.

For comparison, we have shown in Fig.~\ref{b} also the
effective amplitudes $b_1(L_{min})$ and $b_2(L_{min})$
(by "x" and "+", respectively) estimated with the critical exponents of
our GFD theory~\cite{K1,K2} ($\eta=1/8$, $\omega_l=l/2$), assuming
$\delta(t,L)=0$.
The effective amplitudes converge to some values with increasing of
$L_{min}$. These, however, are not the true asymptotic values, since
the maximal size of the system has been eliminated to $L=56$.

\section{A test for the standard 3D Ising model}
\label{stan}

A test with the effective amplitudes, as in Sec.~\ref{imp},
appears to be more sensitive tool as compared to the fits discussed
in Sec.~\ref{sec:fit}. Since more data points are available for the standard
Ising model, we can make even better test than that in Sec.~\ref{imp}.
We have fitted all data points in Tab.~25 of~\cite{HV}
within the interval $[L;8L]$ to the theoretical expression
for $\ln \chi$ (consistent with~(\ref{chi1}))
to evaluate the effective amplitudes $a$ and $b_l$ with $l=1, 2, 3$
depending on $L$. Exceptionally in the case if all the
involved exponents are correct (exact) each effective amplitude can converge
to a certain nonzero asymptotic value at $L \to \infty$. In other words,
if one tries to compensate the inconsistency in the exponent by
choosing appropriate amplitude, then the amplitude
tends either to zero or infinity at $L \to \infty$.

\begin{figure}
\centerline{\psfig{figure=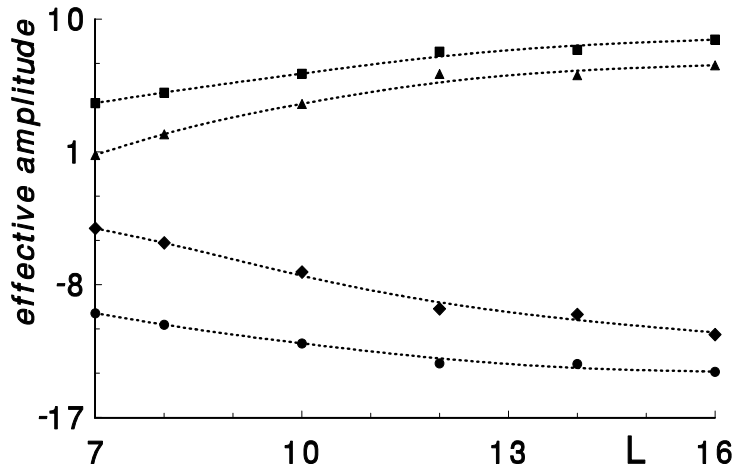,width=11cm,height=8.5cm}}
\vspace{-5ex}
\caption{\small The effective amplitudes in Eq.~(\ref{chi1})
$100 \, (\ln a(L) -1)$ (triangles), $5 b_1(L)$ (circles), $b_2(L)$ (squares),
and $b_3(L)$ (rhombs)
evaluated by fitting the susceptibility data of 3D Ising model at
criticality within the interval of sizes $[L;8L]$ with the critical
exponents $\eta=1/8$ and $\omega_l=l/2$ of the GFD theory.
}
\label{m}
\end{figure}

\begin{figure}
\centerline{\psfig{figure=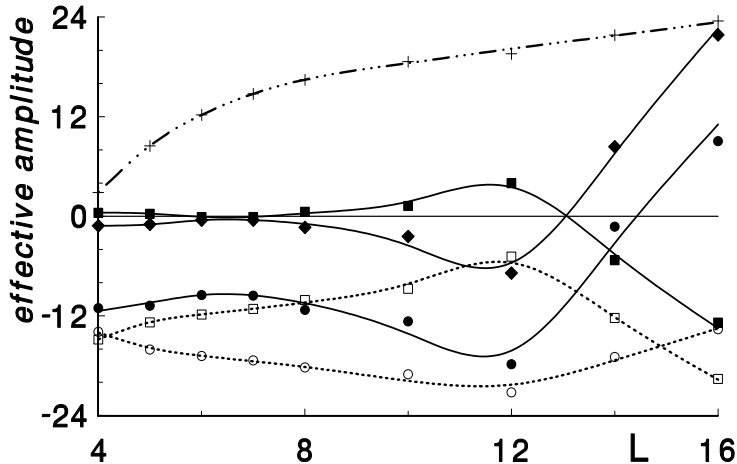,width=11cm,height=8.5cm}}
\vspace{-5ex}
\caption{\small The effective amplitudes in Eq.~(\ref{chi1})
evaluated by fitting the susceptibility data of 3D Ising model at
criticality within the interval of sizes $[L;8L]$ with the critical
exponents $\eta=0.0358$, $\omega_1=0.845$, $\omega_2=2 \omega_1$,
and $\omega_3=2$ proposed in~\cite{Hasenbusch}.
Solid symbols show the four--parameter fit:
$50 b_1(L)$ (circles), $b_2(L)$ (squares), and $b_3(L)$ (rhombs);
empty symbols show the three--parameter fit: $100 b_1(L)$ (circles) and
$27 b_2(L)$ (squares); crosses represent
the amplitude of the two--parameter fit, i.~e., quantity
$190 \, (b_l(L)+0.34)$.
}
\label{v}
\end{figure}

We have shown in Fig.~\ref{m} the effective amplitudes $\ln a(L)$ and
$b_l(L)$ in the case of our critical exponents $\eta=1/8$ and $\omega_l=l/2$.
As we expected, the effective amplitudes  
converge to some nonzero values with increasing of $L$.
This is a good numerical evidence that our critical exponets are true.
The case with the exponents of~\cite{Hasenbusch} $\eta=0.0358(4)$,
$\omega_1=0.845(10)$, $\omega_2=2 \omega_1$, and $\omega_3=2$ 
is illustrated in Fig.~\ref{v}.
As we expected, the effective amplitudes of our four--parameter fit
(solid symbols) tend to diverge with increasing of $L$,
which shows that this set of critical exponents is false.
One could object that, probably, the instability of the effective
amplitudes is due to small errors in MC data. However,
the amplitudes $b_1(L)$ and $b_2(L)$ of the more stable three--parameter
fit ($l=1,2$ in~(\ref{chi1})) behave in a similar way (see empty symbols in
Fig.~\ref{v}). Moreover, the amplitude $b_1(L)$ of the two--parameter fit,
shown by crosses, increases almost
linearly at large enough $L$ instead of the expected (in a case
of correct exponents) saturation like
$b_1(L) \simeq b_1 + const \cdot L^{-\omega}$.
As regards the convergence in Fig.~\ref{m} of the effective amplitudes
at $L \to \infty$, it is possible only if both conditions
are fulfilled, i.~e., the exponents are correct and the MC data
are accurate enough to ensure stable results. Thus, in any case,
the analysis in Fig.~\ref{m} provides rather convincing evidence
that our exponents are the true ones, which by itself rules
out the possibility that those proposed in~\cite{Hasenbusch}
could be correct.
The results in Figs.~\ref{m} and~\ref{v} are affected insignificantly
by a small inaccuracy of about $10^{-7}$ in the estimated $\beta_c$ value.

\section{Some remarks about other numerical results}

There exists a large number of numerical results in the published
literature not discussed here and in our previous papers~\cite{K1,K2}.
A detailed review of these results is given in~\cite{PV}.
The cited there papers report results which disagree with
the values of the critical exponents we have proposed.
However, as regards the pure Monte Carlo study, we are quite confident
that, just like in the actually discussed case of 3D Ising model,
the increase of system syzes and/or use of higher--level
approximations will lead to the conclusion that fits with our
exponents are better than those with the conventional (RG) exponents.
Particularly, a careful analysis of the effective exponents made
in Sec.~\ref{sec:omega} and~\ref{sec:fit}, as well as in Sec.~6
of~\cite{K2} already has shown that the effective exponents deviate from
the values predicted by the perturbative RG theory and converge
more or less to those of the GFD theory at $L \to \infty$.
Together with the analysis of the experiment with superfluid
$^4 He$~\cite{K1}, we have presented totally $6$ independent evidences of
such a behavior.


Formally, the finite--size effects on the obtained values of the
critical exponents has been taken into account in many of cases
considered in literature.
However, the estimated effect strongly depends on that which kind of
corrections to scaling is expected and included in the analysis.
All the existing analysis (not counting our works), of course, are based on
the assumption that the critical behavior of all physical quantities
is characterised by the same correction exponent $\omega$ which is 
about $0.8$ for $O(n)$--symmetric models with $n=1,2,3,4$. However, it is
evident from the behavior of the partition function zeros of
3D Ising model~\cite{K2} that
$\omega$ cannot have so large value. Namely, the value of $(1/\nu)+\omega$
should be about $2$ or even smaller, otherwise we arrive to a rather strong
and obvious contradiction with the MC data for the real part of the partition
function zeros~\cite{K2}. The current analysis in Sec.~\ref{sec:omega}
provides $3$ independent evidences (for $n=1,2,3$) that the correction
exponent for the magnetization cumulant is remarkably smaller than $0.8$.
The numerical analysis often suggests that $\omega \approx 1$. This fact
is perfectly explained by our
theoretical concept: in some cases the amplitude of the leading correction
term can be small as compared to that of the subleading term
providing the effective correction exponent just about $1$.
The value of $\omega$ is crucial for an accurate correction--to--scaling
analysis. If, e.~g., we would assume that $\omega=0.845$, then we could
not arrive to a conclusion that $\eta =1/8$ is a better choice for the
3D Ising model than $\eta=0.0358(4)$, since all fits with $\eta=1/8$ and
$\omega=0.845$ look relatively bad.
This explains the fact that the usual estimations do not give
$\eta \approx 1/8$, while this is just the correct value.
We suppose that similar problems could arise also in other cases,
particularly, if one uses some expression for the correlation length
in finite system (like in~\cite{Ballesteros}), as it has been
discussed in~\cite{K2}.

We should not forget also about purely subjective factor that any
signals about essential inconsistency between MC data and RG predictions
usually are suppressed, i.~e., they do not apper in the published
literature. There are no doubts that such signals exist which can
be mentioned even very easily, e.~g., the behavior of the effective
critical exponents $\omega_{eff}$ discussed in Sec.~\ref{sec:omega},
or those evidences in~\cite{K1} which appear as a result
of unsophisticated analysis of MC data. As a result of
an uncritical acceptance of anything which claims to
confirm with a great accuracy the conventional (RG) values of
the critical exponents and rejection of any contraarguments, the
objective picture is distorted.
This is the reason why almost all the published and reviewed
papers usually claim to confirm with an almost unbelievable
accuracy the predictions of the perturbative RG theory.
It is impossible to check in detail all these papers, but
our critical analysis in~\cite{K1}, \cite{K2}, and here indicates
that many of them are, at least, inobjective.

There exists some background for the conventional claims in the published
literature that all the usual methods give consistent results which appear
to be in a good agreement with the predictions of the perturbative
RG theory. The perturbation expansions of the RG theory, as well
as the techniques of high-- and low--temperature series expansion are
merely not rigorous extrapolation schemes which work not too close to
criticality.
As a result, these methods produce some pseudo or effective critical
exponents which, however, often provide a good approximation
just for the range of temperatures not too close to $T_c$ (critical
temperature) where these methods make sense and, therefore, agree with
each other.
According to the finite--size scaling theory, $t L^{1/\nu}$
is a relevant scaling argument, so that not too small values of the
reduced temperature $t$
are related to not too large sizes $L \sim t^{-\nu}$. Therefore, one
can understand that the MC results for finite systems
often can be well matched to the conventional critical exponents
proposed by high temperature (HT) and RG expansions.
If, however, the level
of MC analysis (i.~e., the level of approximations used) is increased,
then it turns out that the "conventional" critical exponents are
not valid anymore, as it has been demonstrated in the current paper
and in~\cite{K2}. It is because the "conventional" exponents are not
the asymptotic exponents. Correct values of the asymptotic exponents
have been found in~\cite{K1} considering suitable theoretical limits
instead of formal expansions in terms of $\ln k$ (at criticality, where
$k$ is the wave vector magnitude) or $\ln t$ (approaching criticality)
which are meaningless at $k \to 0$ and $t \to 0$. 
These formal expansions lie in the basis of the RG expansions for
the critical exponents. The founders and defenders of the perturbative
RG theory, of course, will try to doubt our statement that the
perturbative RG method is invalid at criticality. But it is impossible
to doubt a mathematical proof. It has been proven in~\cite{K1} that
the assumption that the $\epsilon$--expansion works and provides
correct results at $k \to 0$ leads to an obvious contradiction in
mathematics (cf.~Sec.~2 in~\cite{K1}). This fact alone cannot be
compensated even by an infinite number of numerical evidences
supporting the "conventional" critical exponents coming from the
RG expansions.

Our argument, based on the current numerical analysis, is the
following. We have proposed here a very sensitive method
(i.~e., a study of effective amplitudes) which allows to test the consistency of a given set of
critical exponents with the MC data including several (in our case up to $3$)
corrections to scaling. We have applied this method to one of the recent
and most accurate numerical data for the susceptibility in
3D Ising model, and have got a confirmation that our critical
exponents are true. It would be not correct to doubt our results based on
less sensitive methods and lower--level approximations.

We prefer to rely just on the data of pure MC simulations becose of the
following reasons. The so called Monte Carlo RG (MCRG) method is not
free of assumptions related to approximate renormalization.
We would like only to mention that the MCRG study in~\cite{GT} of 3D Ising
systems of the largest (to our knowledge) available in 
literature sizes, i.~e. up to $L=256$,
has not revealed an excellent agreement with the usual predictions of
the perturbative RG. In particular, an estimate $\omega \approx 0.7$
has been obtained~\cite{GT} which is smaller than the usual (perturbative)
RG value $\approx 0.8$, but still is larger than the exact value $0.5$
predicted by the GFD theory. 
The high--temperature series cannot give more
precise results than those extracted from the recent most accurate MC
data, including the actual data of~\cite{HV}, since these series
diverge approaching the critical point. One approximates the divergent
series by a ratio of two divergent series (Pade approximation),
but it is never proven that such a method converges to the exact result.
It is interesting to compare the MC and HT estimates of the
critical point for the standard 3D Ising model, i.~e.,
$\beta_c \simeq 0.2216545$ (MC)~\cite{HV} and
$\beta_c = 0.221659 +0.000002/-0.000005$ (HT)~\cite{SA}.
It is clear that the MC value is more accurate:
if we look in~\cite{HV}, where the estimation procedure is
well illustrated, we can see that $\beta_c$ is definitely
smaller than $0.221659$, and the error seems to be much
smaller than the difference between both estimates $0.0000045$.
As we have mentioned already, our independent tests suggest
that the error of the actual MC value is about $10^{-7}$.

\newpage

\section{Conclusions}

In summary of the present work, we conclude the following.
\begin{enumerate}
\item
The leading correction--to--scaling exponent $\omega$
in $O(n)$--symmetric models on a three--dimensional
lattice has been tested by analysing the recent Monte Carlo (MC) data.
These tests have shown the incorrectness of some claims that
$\omega$ has a very accurate value $0.845(10)$ at $n=1$.
A selfconsistent infinite volume extrapolation yields
row estimates $\omega \approx 0.547$, $\omega \approx 0.573$, and
$\omega \approx 0.625$ at $n=1$, $2$, and $3$, respectively, in approximate
agreement with the corresponding exact values $1/2$, $5/9$, and $3/5$
predicted by our recently developed GFD theory.
\item
Considering the susceptibility data for 3D Ising model at
criticality~(Sec.~\ref{sec:fit}), we conclude that the fits made with
our (GFD) critical exponents systematically
improve relative to those made with the exponents given in~\cite{Hasenbusch},
as the system sizes grow and the approximation order increases.
\item
The numerical analysis of the effective critical exponents 
in Sec.~\ref{sec:omega} and~\ref{sec:fit}, as well as in Sec.~6
of~\cite{K2} has shown that the effective critical exponents deviate from
the values predicted by the perturbative RG theory and converge
towards those of the GFD theory at $L \to \infty$.
The same behavior has been observed in the experiment with
superfluid $^4 He$ discussed in~\cite{K1}. Totally, these are
$6$ independent evidences of such a behavior, suggesting that the
above examples are not occasional or exceptional,
but reflect a general rule.
\item
Different sets of critical exponents (one provided by GFD theory, another
proposed in~\cite{Hasenbusch}) predicted for the 3D Ising model
have been tested by analysing the effective amplitudes
(Sec.~\ref{imp} and~\ref{stan}).
While the usual fits of the susceptibility data do not allow to show
convincingly which of the discussed here sets of the critical exponents
is better, this method clearly demonstrates that the conventional
critical exponents $\eta=0.0358(4)$ and $\omega=0.845(10)$~\cite{Hasenbusch}
are false, whereas our (GFD) values $\eta=1/8$ and $\omega=1/2$ are true. 
\end{enumerate}

\end{document}